\documentclass{webofc}

\usepackage[varg]{txfonts}   
\usepackage{hyperref}
\usepackage{xcolor}
\usepackage{url}
\hypersetup{colorlinks=true,citecolor=blue,urlcolor=blue,linkcolor=blue}
\begin{document}
\title{Local spin polarization of $\Lambda$ hyperons and its interaction corrections}
%
%

\author{\firstname{Cong} \lastname{Yi}\inst{1,2}
\and
        \firstname{Shuo} \lastname{Fang}\inst{2,3}
        \and
        \firstname{Dong-Lin} \lastname{Wang}\inst{2}
        \and
        \firstname{Shi} \lastname{Pu}\inst{2}\fnsep
        \thanks{{Speaker: shipu@ustc.edu.cn}} 
}

\institute{
Institute of Particle Physics and Key Laboratory of Quark and Lepton
Physics (MOE), Central China Normal University, Wuhan 430079, China  
\and Department of Modern Physics, University of Science and Technology
of China, Hefei, Anhui 230026, China
\and
Technical University of Munich, TUM School of Natural Sciences, Physics Department, James-Franck-Str. 1, 85748 Garching, Germany
          }

\abstract{
We have computed the second Fourier sine coefficient of the longitudinal spin polarization, $\langle P_{z} \sin 2(\phi_{p} - \Psi_{2}) \rangle$, as a function of multiplicity or centrality in Au+Au collisions at $\sqrt{s_{NN}} = 200$ GeV and in $p$+Pb collisions at $\sqrt{s_{NN}} = 8.16$ TeV using the CLVisc hydrodynamic framework. The numerical results successfully describe the data in Au+Au collisions. However, understanding the data in $p$+Pb collisions remains a puzzle. Additionally, we have reported some recent developments in quantum kinetic theory and spin hydrodynamics.
}
\maketitle

\section{Introduction}

In relativistic heavy-ion collisions, two nuclei are accelerated close to the speed of light, accompanied by a huge initial orbital angular momentum on the order of $10^7 \hbar$. Such a large initial orbital angular momentum can lead to the spin polarization of $\Lambda$ hyperons and the spin alignment of vector mesons, as proposed in the early pioneering work~\cite{Liang:2004ph}. The RHIC-STAR collaboration has measured the global polarization of $\Lambda$ hyperons \cite{STAR:2017ckg}. This global polarization can be well described as being induced by the thermal vorticity \cite{Becattini:2017gcx}. 
Subsequently, the polarization along the beam direction as a function of the azimuthal angle $\phi$, referred to as the local polarization of $\Lambda$ hyperons, has also been measured \cite{STAR:2019erd}. It was found that the polarization induced by the shear viscous tensor plays a crucial role in understanding the local polarization, in addition to the contribution from thermal vorticity \cite{Fu:2021pok,Becattini:2021iol,Yi:2021ryh}. For further details, we refer the reader to the recent review \cite{Becattini:2024uha} and references therein.

\begin{figure}
\includegraphics[scale=0.38]{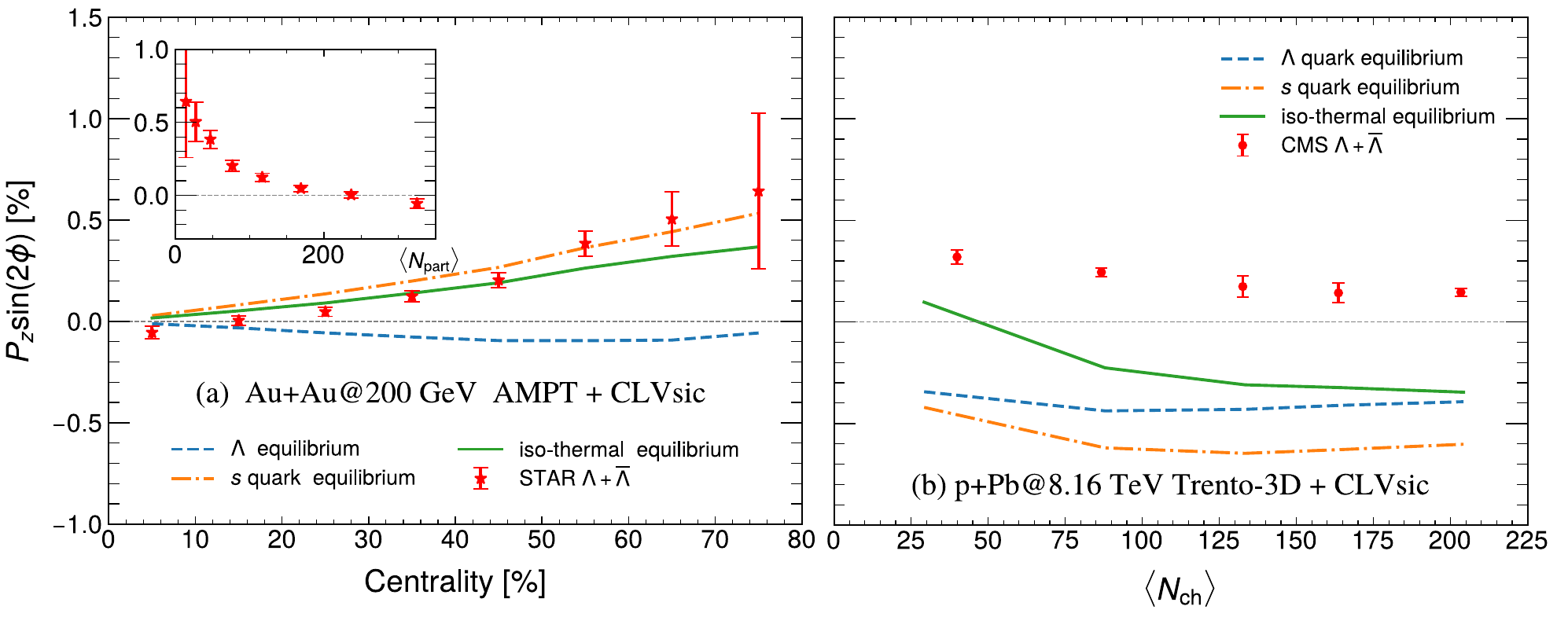}
\caption{The $\langle P_{z}\sin(2\phi_{p}-2\Psi_{2})\rangle$ for $\Lambda$
hyperons as function of centrality or multiplicity: (a) in Au+Au collisions at $\sqrt{s_{NN}}=200$
GeV and (b) in p+Pb collisions at $\sqrt{s_{NN}}=8.16$ TeV. The blue
dashed, orange dash-dotted, and green solid lines represent the results
in the $\Lambda$ equilibrium, $s$-quark equilibrium, and iso-thermal
equilibrium scenarios, respectively. Red markers denote experimental
data from Ref.~\cite{STAR:2019erd} for Au+Au collisions and from Ref.~\cite{CMS:2025nqr} for p+Pb collisions.}
\label{fig:P2z_central} 
\end{figure}

Very recently, the LHC-CMS collaboration measured the second Fourier sine coefficient of the longitudinal spin polarization of $\Lambda$ hyperons, namely $\langle P_{z} \sin 2(\phi_{p} - \Psi_{2}) \rangle$, as a function of centrality or multiplicity in p+Pb collisions at $\sqrt{s_{NN}} = 8.16$ TeV \cite{CMS:2025nqr}, as shown in Fig.~\ref{fig:P2z_central}(b). Surprisingly, the magnitude and trend of $\langle P_{z} \sin 2(\phi_{p} - \Psi_{2}) \rangle$ as a function of multiplicity in p+Pb collisions are very similar to those in Au+Au collisions, as shown in Fig.~\ref{fig:P2z_central}(a). The data suggests a very weak system and collisional energy dependence. It remains unclear how to interpret the results in p+Pb collisions at this stage. 
In this work, we summarize the main results obtained using hydrodynamic simulations and briefly discuss possible corrections arising from interactions between particles. Finally, we also introduce some very recent developments in relativistic spin hydrodynamics.

\section{Centrality dependence of $\langle P_{z} \sin 2(\phi_{p} - \Psi_{2}) \rangle$ in Au+Au and p+Pb collisions}

We implement the (3+1)-D CLVisc hydrodynamic framework to investigate
the local polarization of $\Lambda$ hyperons under three scenarios:
$\Lambda$ equilibrium~\cite{Fu:2021pok,Yi:2021ryh}, $s$-quark
equilibrium~\cite{Fu:2021pok,Yi:2021ryh}, and iso-thermal equilibrium~\cite{Becattini:2021iol}.
The centrality or multiplicity dependence of $\langle P_{z}\sin2(\phi_{p}-\Psi_{2})\rangle$
is presented in Fig.~\ref{fig:P2z_central}(a) for Au+Au collisions
and Fig.~\ref{fig:P2z_central}(b) for p+Pb collisions. The parameters
used in our simulations are tuned to reproduce the observed multiplicity
and elliptic flow. More details about the theoretical framework and
simulation setup can be found in Ref.~\cite{Wu:2022mkr} for Au+Au
collisions and Ref.~\cite{Yi:2024kwu} for p+Pb collisions.

In Fig.~\ref{fig:P2z_central}(a), we observe that our numerical results for 
$\langle P_{z} \sin(2\phi_{p} - 2\Psi_{2}) \rangle$ of $\Lambda$ hyperons in both the $s$-quark equilibrium and iso-thermal equilibrium scenarios agree with the experimental data, whereas the results in the $\Lambda$ equilibrium scenario fail to capture the correct trend. These findings are consistent with those reported in Refs.~\cite{Fu:2021pok, Becattini:2021iol}. 
On the other hand, in Fig.~\ref{fig:P2z_central}(b), we find that our theoretical predictions of $\langle P_{z} \sin(2\phi_{p} - 2\Psi_{2}) \rangle$ for p+Pb collisions under all three scenarios  cannot describe the data. The main reason is that the shear-induced polarization in p+Pb collisions is insufficient to flip the sign of the total local polarization \cite{Yi:2024kwu}. Understanding the local polarization in small systems remains an open puzzle.

\section{Interaction corrections from quantum kinetic theory}

\begin{figure}[t]
\centering
\sidecaption
\includegraphics[scale=0.25]{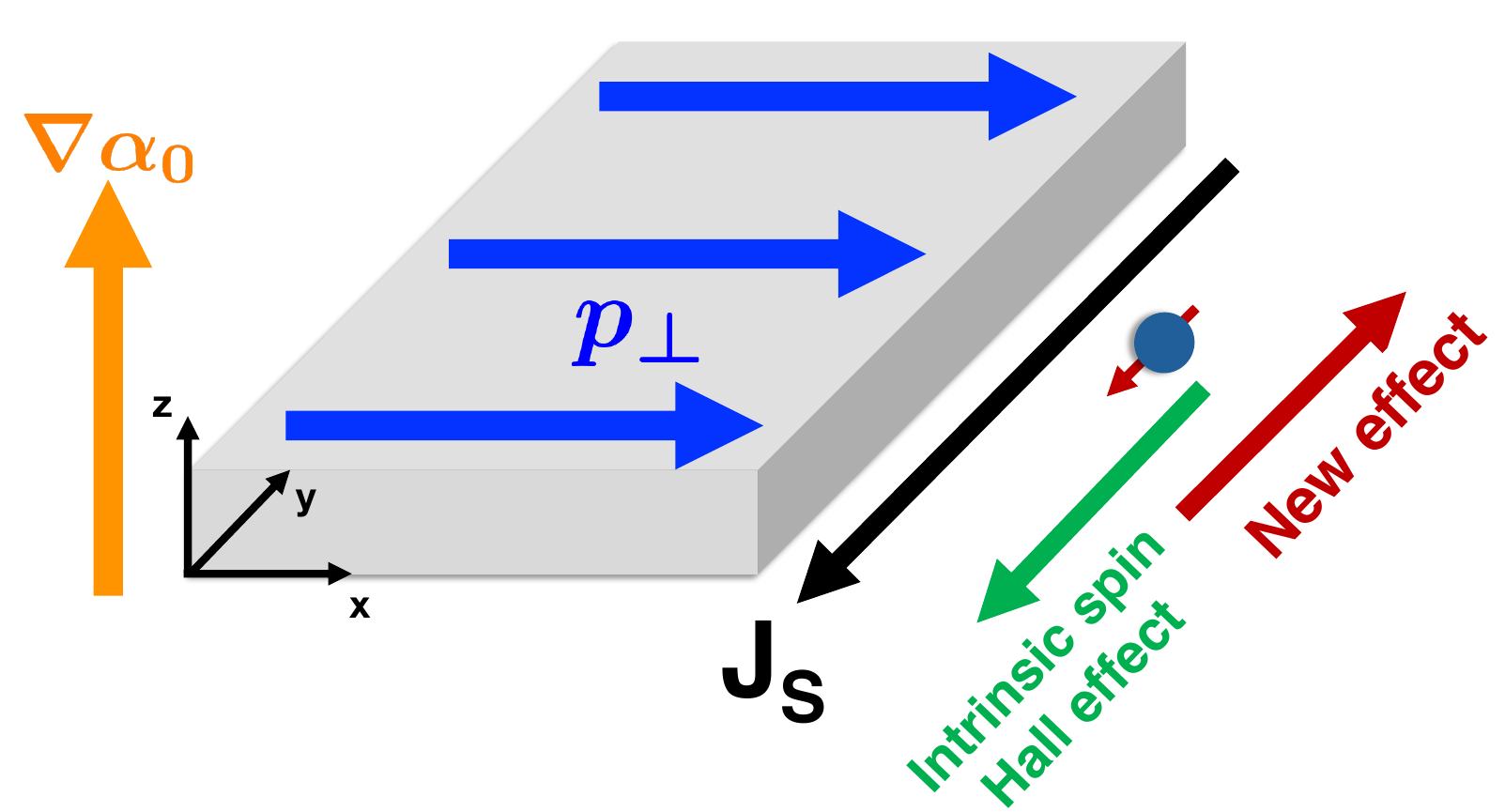}
\caption{
Illustration of the new spin effect induced by interactions. The orange arrow represents the effective field caused by $\nabla \alpha_0$. The blue arrows indicate the direction of the particles' momentum $p_\perp$, which is perpendicular to the effective field. The $J_S$ denotes the direction of polarization, corresponding to $\mathcal{A}^{<,\mu}$ in Eq.~(\ref{eq:SHE}).
}  
\label{fig:New_effect}       
\end{figure}

Though several puzzles regarding the local spin polarization of $\Lambda$ hyperons remain, significant efforts have been made in recent years to investigate off-equilibrium and interaction modifications based on quantum kinetic theory (QKT). QKT provides a systematic framework that incorporates particle and spin transport on the same footing through Wigner functions, rooted in non-equilibrium quantum field theory (for a detailed review, see Ref.~\citep{Hidaka:2022dmn} and references therein). 
The phase-space spin-current density can be expressed in terms of the axial-vector fermion two-point Wigner function, $\mathcal{A}^{<}(q,x)$, under the canonical pseudo-gauge. Currently, there are two primary approaches to incorporate interaction corrections to local polarization. 
The first approach involves directly solving the axial-vector kinetic equation, which includes collisional and background-field effects under quasi-particle approximations. Interaction corrections to the spin current can then be calculated from $\mathcal{O}(\hbar)$ perturbative solutions of the axial-vector Wigner function. 
The second approach coarse-grains QKT into relativistic spin hydrodynamics, where the interaction information is encoded in transport coefficients, such as the spin relaxation time. In this method, off-equilibrium corrections to spin-dependent distribution functions are obtained by solving the spin hydrodynamic equations, which govern the evolution of the moments of these spin-dependent distributions. Consequently, the off-equilibrium modifications to the modified Cooper-Frye formula are derived.


The calculation of the collisional contribution with NJL-type interactions was presented using QKT and the 
method of moments expansion \citep{Weickgenannt:2022zxs, Weickgenannt:2022qvh} under the Hilgevoord-Wouthuysen pseudogauge. 
Subsequently, a resummed spin hydrodynamics was derived using the inverse-Reynolds dominance approach \citep{Wagner:2024fry}, and similar corrections were calculated. These corrections depend on $11$ independent fields, which include the components of the spin potential $\{\omega_{0}^{\mu}, \kappa_{0}^{\mu}\}$ and the spin-stress tensor $\mathfrak{t}^{\mu\nu}$. Numerical simulations of such relativistic spin hydrodynamics, expressed in terms of $\{\omega_{0}^{\mu}, \kappa_{0}^{\mu}, \mathfrak{t}^{\mu\nu}\}$, and their influence on $\Lambda$ spin polarization were presented in Ref.~\citep{Sapna:2025yss}.


An alternative calculation of the off-equilibrium corrections was performed by directly solving quantum kinetic equations, including solving  vector and axial-vector distribution functions, $f_{{\rm V/A}}$, as presented in Ref.~\citep{Fang:2024vds}. The leading gradient contribution originates from the collisional term $\mathcal{C}_{{\rm V},\nu}[\delta f_{{\rm V}}]$, with $\delta f_{{\rm V/A}} = f_{{\rm V/A}} - f_{{\rm V/A,leq}}$, and results in a cancellation of the coupling constant. For instance, in the QED-type $2$-to-$2$ scattering process under the hard-thermal-loop approximation, this contribution is expressed as,
\begin{eqnarray}
\delta\mathcal{A}_{\text{SHE,sj}}^{<,\mu}(p) & = & -2\pi\hbar\theta(p_{0})\delta(p^{2})\beta_{0}\frac{\epsilon^{\mu\nu\rho\sigma}p_{\rho}u_{\sigma}}{2E_{\mathbf{p}}}\left[g_{1}(E_{\mathbf{p}})\nabla_{\nu}\alpha_{0} + g_{2}(E_{\mathbf{p}})\sigma_{\nu\alpha}p^{\alpha}\right], \label{eq:SHE}
\end{eqnarray}
where $\beta_0$ is the inverse temperature, $g_{1,2}$ are functions of the particle's energy $E_{\mathbf{p}}$, $\alpha_0 = \beta_0 \mu$ with $\mu$ being the chemical potential, and $\sigma_{\mu\nu}$ is the shear viscous tensor. 
Eq.~(\ref{eq:SHE}) captures the side-jump mechanism in the phase-space spin Hall effect, which results from the accumulation of jump shifts during each collision event and can be traced back to Berry curvature effects and is illustrated in Fig.~\ref{fig:New_effect}. The explicit expressions for $g_{1,2}$ and additional relevant works can be found in Ref.~\citep{Fang:2024vds}.
On the other hand, the skew-scattering contribution, arising from spin-dependent (non-local) collisions via $\delta f_{{\rm A}}$, is of the order $\mathcal{O}\left((g^{4}\ln g^{-1})\times\partial^{2}\right)$ \citep{Fang:2024vds}. 
In a weakly coupled quark-gluon plasma, another contribution arises from medium semi-long-wavelength excitations of $\lambda \sim (gT)^{-1}$. Here, the spacetime gradient of the in-medium quark self-energy, $\partial_{\mu}\Sigma_{{\rm V},\nu} \sim F_{\mu\nu}$, acts as an effective background electromagnetic field that serves as a source of spin torque and induces Hall-type currents \citep{Fang:2023bbw}.
Nevertheless, it remains an open question how to investigate spin polarization in a strongly coupled QCD plasma by generalizing QKT beyond the quasi-particle regime.

\section{Some developments on spin hydrodynamics}

The macroscopic approach to describe spin evolution in relativistic heavy-ion collisions is to add the spin degree of freedom to relativistic hydrodynamics, leading to the framework of relativistic spin hydrodynamics \cite{Florkowski:2018fap, Hattori:2019lfp, Fukushima:2020ucl}. The basic idea in canonical spin hydrodynamics is to decompose the rank-3 total angular momentum tensor as,
$J^{\lambda\mu\nu} = x^{\mu}T^{\lambda\nu} - x^{\nu}T^{\lambda\mu} + \Sigma^{\lambda\mu\nu}, $
where $T^{\mu\nu}$ and $\Sigma^{\lambda\mu\nu}$ represent the energy-momentum tensor and spin tensor, respectively. The first two terms in the decomposition constitute the orbital angular momentum. The evolution equation for the spin tensor follows from the decomposition of the total angular momentum and the conservation of energy, momentum, and total angular momentum, $\partial_{\lambda}\Sigma^{\lambda\mu\nu} = -2T^{[\mu\nu]}$,
where the antisymmetric part of the energy-momentum tensor, $T^{[\mu\nu]} \equiv (T^{\mu\nu} - T^{\nu\mu})/2$, serves as a spin torque that describes the transfer of angular momentum between orbital and spin components. 
For more details, we refer to recent reviews in Refs.~\cite{Becattini:2024uha, Shi:2023sxh} and references therein. 
Recently, spin hydrodynamics with a totally antisymmetric spin tensor $\partial_{\lambda}\Sigma^{\lambda\mu\nu}$ has been proposed in Ref.~\cite{Fang:2025aig}. The resulting evolution equation for spin density incorporates the effects of Thomas precession and couplings between spin and hydrodynamic motion, such as rotation and expansion, thereby generalizing the well-known Bargmann-Michel-Telegdi equation to hydrodynamics.

Despite significant progress in recent years, several challenges remain in the development of spin hydrodynamics. One major issue is the choice of pseudo-gauges. Different pseudo-gauges lead to different definitions of the spin tensor and, consequently, distinct evolution equations \cite{Speranza:2020ilk}. For instance, in the canonical gauge, $T^{[\mu\nu]} \neq 0$, and spin tensor is not conserved \cite{Hattori:2019lfp, Fukushima:2020ucl}, whereas in the Hilgevoord-Wouthuysen gauge, $T^{[\mu\nu]} = 0$, and spin tensor is conserved \cite{Weickgenannt:2022zxs}. Notably, the spin tensor vanishes in the Belinfante gauge \cite{Fukushima:2020ucl}. 
Presently, there is no consensus on which gauge is physically preferred (see also the recent discussions in Refs.~\cite{Buzzegoli:2021wlg, Buzzegoli:2024mra, Becattini:2025twu}).
Another issue lies in the thermodynamic relationships commonly used in spin hydrodynamics, which have been shown to be incomplete. More comprehensive versions have been proposed using quantum statistical methods \cite{Becattini:2023ouz, Becattini:2025oyi} and kinetic theory \cite{Florkowski:2024bfw}, but further investigations are required to understand the implications of new corrections. 
Additionally, spin hydrodynamic equations are known to exhibit 
acausal modes \cite{Xie:2023gbo} and unstable modes \cite{Daher:2022wzf}. Acausal modes can be eliminated by properly including relaxation times, while the unstable modes require distinguishing spin susceptibilities for the electric and magnetic components of the spin density. However, this approach lacks a clear physical interpretation and requires further exploration.

\section{Summary}
The observable $\langle P_{z} \sin 2(\phi_{p} - \Psi_{2}) \rangle$, as a function of centrality or multiplicity in high-energy nucleus-nucleus collisions, has been successfully described by hydrodynamic models. However, these models fail to reproduce the behavior of $\langle P_{z} \sin 2(\phi_{p} - \Psi_{2}) \rangle$ in $p$+Pb collisions.
There have been significant advancements in both microscopic and macroscopic approaches. Within the framework of quantum kinetic theory, various interaction corrections have been computed, contributing to local spin polarization. On the other hand, in spin hydrodynamics, although substantial progress has been made, the pseudo-gauge dependence remains an unresolved puzzle.

\section*{Acknowledgements:} 
This work is supported in part by the National Key Research
and Development Program of China under Contract No. 2022YFA1605500,
by the Chinese Academy of Sciences (CAS) under Grant No. YSBR-088.

\bibliography{qkt-ref}

\end{document}